\documentclass[twocolumn,showpacs,preprintnumbers,amsmath,amssymb]{revtex4}

\usepackage{epsfig}
\usepackage{graphicx}
\usepackage{dcolumn}
\usepackage{bm}

\def\DESepsf(#1 width #2){\epsfxsize=#2 \epsfbox{#1}}

\begin{document}


\title{\boldmath{Indication for Large Rescatterings in
Charmless Rare $B$ Decays}}

\author{$^{a)}$Chun-Khiang Chua}
\author{$^{a)}$Wei-Shu Hou}%
\author{$^{b)}$Kwei-Chou Yang}
\affiliation{%
$^{a)}$Department of Physics, National Taiwan University, Taipei,
Taiwan 10764, Republic of China\\
$^{b)}$Department of Physics, Chung Yuan Christian University,
Chung-Li, Taiwan 32023, Republic of China
}%

\date{\today}

\begin{abstract}
The current wealth of charmless $B$ decay data may suggest the
presence of final state rescattering. In a factorized amplitude
approach, better fits are found by incorporating two SU(3)
rescattering phase differences, giving $\delta \sim 65^\circ$ and
$\sigma \sim 90^\circ$--$100^\circ$. Fitting with unitarity phase
$\phi_3$ as a fit parameter gives $\phi_3 \sim 96^\circ$, the $CP$
asymmetries $A_{\pi\pi}$, $S_{\pi\pi}$ agree better with BaBar,
and the $\sigma$ phase is slightly lower. Keeping $\phi_3 =
60^\circ$ fixed in fit gives $S_{\pi\pi}\sim -0.9$, which agrees
better with Belle. With the sizable $\delta$, $\sigma$
rescattering phases as fitted, many direct $CP$ asymmetries flip
sign, and $B^0\to \pi^0\pi^0$, $K^-K^+$ rates are of order
$10^{-6}$, which can be tested soon.
\end{abstract}

\maketitle

\section{Introduction}

Based on data from the CLEO experiment, it was pointed out in
1999~\cite{gamma} that the emerging $B\to K\pi$, $\pi\pi$ rates
support factorization, {\it if the phase angle $\gamma$}
 (or $\phi_3 \equiv \arg V_{ub}^*$~\cite{PDG}) of
the Cabibbo-Kobayashi-Maskawa (CKM) quark mixing matrix {\it is
large}. We now have some theoretical basis for
factorization~\cite{QCD,PQCD} in charmless $B$ decays, and it is
common for $B$ physics practitioners to take $\phi_3 \sim
80^\circ$--$90^\circ$, which is in contrast with the $\sim
60^\circ$ value indicated by CKM fits~\cite{CKM} to other data.

At the turn of the century,
the dramatic ascent of the B
factories brought the CLEO era to an end. 
CLEO did make an attempt~\cite{CLEO_CP} to measure direct $CP$
violating rate asymmetries ($A_{\rm CP}$) in a few modes, where
for a $B\to f$ decay
\begin{equation}
A_{\rm CP}(f)\equiv\frac{{\mathcal B}(\overline B\to\overline f)-{\mathcal B}(B\to f)}
               {{\mathcal B}(\overline B\to\overline f)+{\mathcal B}(B\to f)}.
\end{equation}
The central values of the CLEO measurements differed from
factorization expectations, but errors were large. Some other
oddities lingered, such as the smallness of $\pi^-\pi^+$ rate
compared with $\pi^-\pi^0$. It was thus suggested~\cite{delta}
that one {\it may need large rescattering, in addition to large
$\phi_3$}. Though speculative, this had implications beyond the
shifted pattern in $A_{\rm CP}$s: $\pi^0\pi^0$ would become
prominent, and $CP$ violation in $\overline B{}^0\to \pi^-\pi^+$
could be sizable. A few years later, it is surprising that the
speculated patterns seem to be really emerging!

Let us give a snapshot of the current
landscape~\cite{hh_belle,hh_babar,hh_cleo,pipi_belle,pipi_babar,moriond}
(as summarized in Tables~\ref{tab:rate} and \ref{tab:ACPs}). All
the $K\pi$, $\pi\pi$ modes have been measured by both Belle and
BaBar experiments to some accuracy, and with good agreement. The
$A_{\rm CP}$ in $K^-\pi^{+}$ mode is now significant, and seem
{\it opposite in sign} to the factorization
expectation~\cite{QCD}; the 3$\sigma$ effect in $\overline
K{}^0\pi^-$ mode reported by Belle is no more. The conflicting
signs between Belle and BaBar on $A_{\rm CP}$ in $K^-\pi^{0}$ and
$\overline K^0\pi^{-}$ modes keeps these asymmetries rather
consistent with zero. Both experiments have hints for
$\pi^0\pi^0\sim 10^{-6}$, and have measured the mixing dependent
$CP$ asymmetries in $\pi^-\pi^+$ mode. For the latter, the two
experiments diverge. We thus scale the error-bars by an $S$ factor
following the Particle Data Group (PDG)~\cite{PDG} for these
modes.
Following Belle, we denote the two measurables as $A_{\pi\pi}$
($\equiv -C_{\pi\pi}$), equivalent to $A_{\rm CP}$ in $\overline
B{}^0\to \pi^-\pi^+$, and $S_{\pi\pi}$, where
\begin{eqnarray}
A^{\pi\pi}_{\rm CP}(t)&=&\frac{{\mathcal B}(\overline B {}^0(t)\to\pi^+\pi^-)-
                      {\mathcal B}(B^0(t)\to \pi^+\pi^-)}
               {{\mathcal B}(\overline B {}^0(t)\to\pi^+\pi^-)+
                      {\mathcal B}(B^0(t)\to \pi^+\pi^-)}
\nonumber\\
&=&A_{\pi\pi}\cos(\Delta m_B t)+S_{\pi\pi} \sin(\Delta m_B t).
\end{eqnarray}
Belle finds $A_{\pi\pi}\simeq 1$ and $S_{\pi\pi}\simeq
-1$~\cite{pipi_belle}
 (hence outside the physical domain), while
BaBar finds~\cite{pipi_babar} both $\simeq 0$.
Note that Belle's updated numbers
are consistent with earlier results,
and the discrepancy with BaBar~\cite{pipi_babar} remains.
The two central values for $A_{\pi\pi}$ now do agree in sign,
but are {\it opposite} to (QCD) factorization expectations~\cite{QCD}.

Independent hint for rescattering~\cite{CHY} comes from the
unexpected emergence of color-suppressed $\overline B{}^0\to
D^{(*)0}h^0$ decays~\cite{D0h0}, where $h^0 = \pi^0$, $\eta$ and
$\omega$; the rates are all larger than expected. In this paper we
extend the model of Ref.~\cite{delta} and explore the implications of
present data on both $\phi_3$ and rescattering phases.
The pattern change in rates, such as $\pi^0\pi^0$, $K^-K^{+,0}$,
the sizable ``{\it opposite sign}" $A_{\rm CP}$s in various modes,
and especially $A_{\pi\pi}$ and $S_{\pi\pi}$, can
be tested in the near future.

\begin{table*}[hbt]
\begin{center}
\caption{Experimental data for CP-averaged $\overline B\to hh$
branching ratios in units of
$10^{-6}$~\cite{hh_belle,hh_babar,hh_cleo,pipi_belle,pipi_babar,moriond}.
\label{tab:rate}}
\begin{tabular}{lcccr}\hline \hline
${\cal B}\times 10^6$
        & CLEO
        & BaBar
        & Belle 
        & Average  \\
\hline
$K^-\pi^+  $
        & $18.0^{+2.3}_{-2.1}{}^{+1.2}_{-0.9}$
        & $17.9 \pm 0.9\pm 0.7$
        & $18.5 \pm 1.0 \pm 0.7$
        & $18.2\pm0.8$
\\
$K^- \pi^0$
        & $12.9^{+2.4}_{-2.2}{}^{+1.2}_{-1.1}$
        & $12.8^{+1.2}_{-1.1}\pm 1.0$
        & $12.8\pm1.4^{+1.4}_{-1.0}$
        & $12.8 \pm 1.1$
\\
$\overline K {}^0 \pi^-$
        & $18.8^{+3.7}_{-3.3}{}^{+2.1}_{-1.8}$
        & $20.0\pm1.6\pm1.0$
        & $22.0\pm1.9\pm 1.1$
        & $20.6\pm 1.3$
\\
$\overline K {}^0 \pi^0$
        & $12.8^{+4.0}_{-3.3}{}^{+1.7}_{-1.4}$
        & $10.4 \pm 1.5 \pm 0.8$
        & $12.6\pm2.4\pm 1.4$
        & $11.2 \pm 1.4 $
\\
$\pi^- \pi^0$
        & $4.6^{+1.8}_{-1.6}{}^{+0.6}_{-0.7}$
        & $5.5^{+1.0}_{-0.9}\pm 0.6$
        & $5.3\pm1.3\pm 0.5$
        & $5.3 \pm 0.6 $
\\
$\pi^+ \pi^-$
        & $4.5^{+1.4}_{-1.2}{}^{+0.5}_{-0.4}$
        & $4.7 \pm 0.6 \pm 0.2$
        & $4.4 \pm 0.6\pm 0.3$
        & $4.5 \pm 0.4$
\\
$\pi^0 \pi^0$
        & $2.2^{+1.7}_{-1.3}\pm 0.7$
        & $1.6^{+0.7}_{-0.6}{}^{+0.6}_{-0.3}$
        & $1.8^{+1.4}_{-1.3}{}^{+0.5}_{-0.7}$
        & $1.7 \pm 0.6$
\\
$\eta \pi^- $
        & ---
        & $4.2\pm1.0\pm0.3$
        & $5.4^{+2.0}_{-1.7}\pm0.6$
        & $3.9\pm 0.8$
\\
$\eta K^- $
        & ---
        & $2.8\pm0.8\pm0.2$
        & $ 5.3^{+1.8}_{-1.5}\pm 0.6 $
        & $3.2\pm 0.7$
\\
$\eta \overline K {}^0 $
        & ---
        & $<4.6$
        & ---
        & $<4.6$
\\
$K^-  K^0$
        & $<3.3$
        & $<2.2$
        & $(1.7\pm1.2\pm0.1)<3.4$
        & $<2.2$
\\
$K^- K^+  $
        & $<0.8$
        & $<0.6$
        & $<0.7$
        & $<0.6$
\\
$\overline K^0 K^0$
        & $<3.3$
        & $<1.6$
        & $(0.8\pm0.8\pm0.1)<3.2$
        & $<1.6$
\\
\hline \hline
\end{tabular}
\end{center}
\end{table*}

\begin{table*}[t]
\vspace{0.3truecm}
\begin{center}
\caption{Experimental data for direct CP asymmetries in $\overline
B\to hh$
decays~\cite{CLEO_CP,hh_babar,pipi_belle,pipi_babar,moriond}.
\label{tab:ACPs}}
\begin{tabular}{lcccr}
\hline \hline
$A_{\rm CP}$~(\%)
        & CLEO 
        & BaBar 
        & Belle 
        & Average
\\
\hline
$K^-\pi^+ $
        & $-4\pm 16$
        & $-10.2\pm 5\pm1.2$
        & $-7\pm 6\pm 1$
        & $-9\pm4$
\\
$K^-\pi^0 $
        & $-29\pm 23$
        & $-9\pm 9\pm 1$
        & $23\pm 11^{+1}_{-4}$
        & $1\pm 12$\footnotemark[1]
\\
$\overline K^0\pi^-$
        & $18\pm 24$
        & $-5.3\pm 7.9\pm1.3$
        & $7^{+9}_{-8}{}^{+1}_{-3}$
        & $1\pm 6$
\\
$\overline K {}^0\pi^0 $
        & ---
        & $3\pm 36\pm 9$
        & ---
        & $3\pm 36$
\\
$\pi^-\pi^0$
        & ---
        & $-3\pm 18\pm 2$
        & $-14\pm 24^{+5}_{-4}$
        & $-7\pm14$
\\
$A_{\pi\pi}$
        & ---
        & $30\pm25\pm4$
        & $77\pm27\pm8$
        & $51\pm23$\footnotemark[2] 
\\
$S_{\pi\pi}$
        & ---
        & $2\pm34\pm5$
        & $-123\pm41^{+8}_{-7}$
        & $-49\pm61$\footnotemark[2] 
\\
$\eta K^-$
        & ---
        & $-32\pm 20\pm 1$
        & ---
        & $-32\pm20$
\\
$\eta \pi^-$
        & ---
        & $-51\pm 19\pm 1$
        & ---
        & $-51\pm19$
\\
\hline\hline
\end{tabular}
\end{center}
\footnotetext[1]{Error-bars of $A^{\rm expt}_{\rm CP}(K^-\pi^0)$
is scaled by $S=1.8$.}
\footnotetext[2] {Error-bars of $A_{\pi\pi}$, $S_{\pi\pi}$ are
scaled by $S=1.2$, 2.3, respectively.}
\end{table*}

\section{Formalism}

Our picture is that of factorized $B$ decay amplitudes followed by
{\it final state} rescattering (FSI), i.e.
\begin{equation}
\langle i; \mbox{\rm out}|H_{\rm W}|B\rangle
 = \sum_l {\cal S}^{1/2}_{il} {\cal A}^f_{l},
\label{eq:master}
\end{equation}
where $| i; \mbox{\rm out}\rangle$ is the out state, ${\cal S}$ is
the strong scattering matrix, and ${\cal A}^f_{l}$ is a
factorization amplitude.
It was pointed out in Ref.~\cite{Donoghue} that elastic
rescattering effects may not yet be greatly suppressed at $m_B$
scale, while inelastic rescattering contributions may be
important.

We would clearly lose control if the full structure of
Eq.~(\ref{eq:master}) is employed. It is clear that the subset of
two body final states from elastic rescatterings stand out against
inelastic channels. Furthermore, it has been shown from
duality~\cite{GH} as well as statistical arguments~\cite{SW} that
inelastic FSI amplitudes tend to cancel among themselves and lead
to small FSI phases $\lesssim 20^\circ$. Thus, quasi-elastic
scattering may well be still relevant in two-body $B$ decays, and
we should allow experiment as the final judge.
For example, if the factorization amplitudes are already good
enough, the data will force us to have ${\cal S}^{1/2}\sim 1$, and
vice versa.

\begin{table*}[t!]
\caption{ World average inputs and fitted outputs; data in
brackets are not used in fit, while $\eta_8 K(\pi)^-$ entries are
for $\eta K(\pi)^-$. Horizontal lines separate rescattering
subsets. Fit 1 or 2 stand for $\phi_3$ free or fixed at
$60^\circ$. Setting $\delta = \sigma = 0$ but keeping other
parameters fixed give the results in parentheses; the fitted
parameters and $\chi^2_{\rm min.}$ are given in
Table~\ref{tab:phase}.
 \label{tab:output} }
\begin{ruledtabular}
\begin{tabular}{lcrrcrr}
 Modes
 & ${\mathcal B}^{\rm expt}\times10^6$
 & ${\mathcal B}^{\rm Fit1}\times10^6$
 & ${\mathcal B}^{\rm Fit2}\times10^6$
 & $A_{\rm CP}^{\rm expt}~(\%)$
 & $A_{\rm CP}^{\rm Fit1}~(\%)$
 & $A_{\rm CP}^{\rm Fit2}~(\%)$\\
\hline

$K^-\pi^+$
        & $18.2\pm0.8$
        & $19.4^{+1.0}_{-1.2}$ $(19.7)$
        & $18.5\pm0.6$ $(18.1)$
        & $-9\pm 4$
        & $-6^{+2}_{-3}$ $(9)$
        & $-4\pm 1$ $(7)$\\
$\overline K {}^0\pi^0$
        & $11.2\pm1.4$
        & $8.3^{+1.3}_{-0.4}$ $(7.3)$
        & $9.1\pm 0.3$ $(8.7)$
        & $\,\,\,\,[3\pm 37]$
        & $24^{+4}_{-30}$ $(0)$
        & $16^{+2}_{-21}$ $(0)$ \\
$\overline K {}^0 \eta_8$
        & [$< 4.6$ (90\% CL)]
        & $3.4^{+0.8}_{-0.6}$ $(4.1)$
        & $3.9^{+1.0}_{-0.8}$ $(4.6)$
        & ---
        & $24^{+9}_{-4}$ $(0)$
        & $15^{+3}_{-2}$ $(0)$\\
\hline $\overline K {}^0 \pi^-$
        & $20.6\pm1.3$
        & $19.6^{+2.2}_{-1.4}$ $(18.5)$
        & $21.6\pm 0.6$ $(20.9)$
        & $1\pm 6$
        & $8^{+2}_{-1}$ $(0)$
        & $5\pm 0$ $(0)$\\
$K^-\pi^0$
        & $12.8\pm1.1$
        & $11.6^{+0.5}_{-1.0}$ $(12.1)$
        & $11.0\pm 0.3$ $(10.9)$
        & $1\pm 12$
        & $-19^{+4}_{-7}$ $(7)$
        & $-14^{+1}_{-2}$ $(6)$\\
$K^-\eta_8$
        & $[3.2\pm0.7]$
        & $3.6^{+0.8}_{-0.7}$ $(4.2)$
        & $4.6^{+1.0}_{-0.8}$ $(5.4)$
        & $[-32\pm20]$
        & $\,\,\,\, 33^{+15}_{-9}$ $(-9)$
        & $\,\,\,\, 19^{+6}_{-4}$ $(-5)$\\
\hline $\pi^-\pi^0$
        & $5.3\pm0.6$
        & $4.4^{+1.2}_{-0.6}$ $(4.4)$
        & $3.2^{+0.1}_{-0.2}$ $(3.2)$
        & $\,\,\,\,[-7\pm 14]$
        & $0$ $(0)$
        & $0$ $(0)$ \\
\hline $\pi^-\eta_8$
        & $[3.9\pm0.8]$
        & $1.2^{+0.1}_{-0.3}$ $(1.4)$
        & $1.5^{+0.0}_{-0.1}$ $(1.8)$
        & $[-51\pm19]$
        & $\,\,\,\, 75^{+25}_{-18}$ $(-32)$
        & $\,\,\,\, 42^{+9}_{-6}$ $(-19)$\\
$K^-K^0$
        & $< 2.2$ (90\% CL)
        & $1.7^{+0.3}_{-0.2}$ $(1.5)$
        & $1.3\pm 0.1$ $(1.0)$
        & ---
        & $-84^{+9}_{-14}$ $(-4)$
        & $-79\pm3$ $(-3)$\\
\hline $\pi^- \pi^+$
        & $4.5\pm0.4$
        & $4.7^{+0.7}_{-0.8}$ $(7.4)$
        & $5.1^{+0.3}_{-0.4}$ $(8.7)$
        & $[A_{\pi\pi}=\,\,\,\,51\pm 23]$
        & $A_{\pi\pi}=12^{+14}_{-50}$ $(-24)$
        & $A_{\pi\pi}=\,\,\,\,13\pm 2$ $(-17)$\\
        &
        &
        &
        & $[S_{\pi\pi}=-49\pm 61]$
        & $S_{\pi\pi}= -15^{+69}_{-0}$ $(-5)$
        & $S_{\pi\pi}=-90^{+1}_{-0}$ $(-88)$\\
$\pi^0\pi^0$
        & $1.7\pm 0.6$
        & $2.5^{+0.4}_{-0.9}$ $(0.2)$
        & $2.8^{+0.3}_{-0.7}$ $(0.1)$
        & ---
        & $-56^{+1}_{-16}$ $(1)$
        & $-35^{+1}_{-6}$ $(1)$\\
$K^-K^+$
        & $<0.6$ (90\% CL)
        & $0.2^{+0.4}_{-0.2}$ $(0.0)$
        & $0.6^{+0}_{-0.1}$ $(0.0)$
        & ---
        & $-13^{+9}_{-76}$ ( --- )
        & $-11^{+2}_{-1}$ ( --- )\\
$\overline K {}^0K^0$
        & $<1.6$ (90\% CL)
        & $1.5^{+0.3}_{-0.6}$ $(1.5)$
        & $1.1\pm 0.1$ $(1.0)$
        & ---
        & $-63^{+141}_{-24}$ $(-4)$
        & $-86^{+6}_{-1}$ $(-3)$\\
$\pi^0\eta_8$
        & ---
        & $0.4^{+0.2}_{-0}$ $(0.3)$
        & $0.2\pm0.0$ $(0.2)$
        & ---
        & $-3\pm0$ $(-4)$
        & $-3\pm0$ $(-4)$\\
$\eta_8\eta_8$
        & ---
        & $0.2\pm0.0$ $(0.1)$
        & $0.2\pm0.1$ $(0.1)$
        & ---
        & $-10^{+96}_{-75}$ $(-6)$
        & $-91^{+14}_{-3}$ $(-6)$\\
\end{tabular}
\end{ruledtabular}
\end{table*}

We extend from the quasi-elastic $\overline B\to DP$~\cite{CHY}
case to the $\overline B\to PP$ case, where $D$ is the SU$(3)$
$D$-meson triplet and $P$ the pseudoscalar octet. That is, we
extend the ${\bf 3} \bigotimes {\bf 8} \to {\bf 3} \bigotimes {\bf
8}$ FSI formalism developed for color-suppressed $D^0h^0$ modes,
to ${\bf 8} \bigotimes {\bf 8} \to {\bf 8} \bigotimes {\bf 8}$
rescattering in light $PP$ final states. Bose symmetry then
implies that the ${\cal S}^{1/2}$ matrix in Eq.~(\ref{eq:master})
takes up the form
\begin{equation}
{\cal S}^{1/2}=
 e^{i\delta_{\bf 27}} |{\bf 27}\rangle\langle {\bf 27}|
+e^{i\delta_{\bf  8}} |{\bf  8}\rangle\langle {\bf  8}|
+e^{i\delta_{\bf  1}} |{\bf  1}\rangle\langle {\bf  1}|,
\label{eq:Smatrix}
\end{equation}
hence there are just two physical phase differences, which we take
as $\delta \equiv \delta_{\bf 27} - \delta_{\bf 8}$ and $\sigma
\equiv \delta_{\bf 27} - \delta_{\bf 1}$. These rescattering
phases redistribute the factorized decay amplitudes ${\cal
A}^f_{l}$ according to Eq. (\ref{eq:master}),
but they also can be viewed as {\it a simple two parameter model
extension} beyond the usual $B\to PP$ amplitudes.
The detailed formalism is given elsewhere. There are also some
similar works in the literature~\cite{similar}. We shall see that
$\vert \delta - \sigma \vert \lesssim 50^\circ$ turns out to be
the case of interest.
As in~\cite{CHY}, we drop the SU$(3)$ singlet ${\bf 1}$, or
$\eta_1$, from the rescattering formulation. For the present case,
one anyway has the difficulty in explaining the huge rate
observed~\cite{PDG} for $B\to \eta^\prime K$.

For ${\cal A}^f_{l}$, the present QCD~\cite{QCD} or
PQCD~\cite{PQCD} factorization approaches involve effects that are
subleading in $1/m_b$, such as weak annihilation contributions. To
avoid double counting hadronic effects, we shall
use~\cite{absorpt} naive factorization~\cite{AKL} amplitudes. In
fact, for QCD factorization, which is close to naive
factorization, removing all subleading effects but keeping the
$1/m_s^{\rm eff}$ chiral enhancement seems to give a better fit to
$K\pi$, $\pi\pi$ rates~\cite{Beneke}.

We follow the $\chi^2$ fit strategy of Ref.~\cite{HSW},
but keeping the $\delta$ and $\sigma$ phases of
Eqs. (\ref{eq:master}) and~(\ref{eq:Smatrix}) as additional parameters.
As input, we take central values of $\vert V_{cb}\vert$
 and $\vert V_{ub}\vert$ from Ref.~\cite{PDG}.
We focus on $\overline B\to PP$ modes, since the $VP$, $VV$
situation is not yet settled. We thus take only the 7 $K\pi$ and
$\pi\pi$ rates, the better measured $A_{\rm CP}$s
 in $K^-\pi^{+,0}$ and $K^0\pi^-$ modes,
and average over the current values reported by BaBar, Belle and
CLEO~\cite{CLEO_CP,hh_belle,hh_babar,hh_cleo,pipi_belle,pipi_babar,moriond},
as given in Table~\ref{tab:rate} and \ref{tab:ACPs}.
We also constrain $f_\pi F_0^{BK}/f_K F_0^{B\pi}=0.9\pm
0.1$~\cite{QCD}. We use the stringent upper limit~\cite{moriond}
of $2.2\; (0.6) \times 10^{-6}$ for $K^-K^{0\,(+)}$ of BaBar. The
limits are implemented as boundaries. {\it We do not use
$A_{\pi\pi}$ and $S_{\pi\pi}$ as input}, since the experimental
situation is still controversial. We do not use $\eta\pi^-(K^-)$
rates and $A_{\rm CP}$s, but give $\eta_8\pi^-(K^-)$ only.

The fit parameters are $F_0^{BK}$, the chiral enhancement
parameter $1/m_s^{\rm eff}$~\cite{gamma}, the FSI phases
$\delta$ and $\sigma$, and possibly $\phi_3$. Because of the
$S_{\pi\pi}$ controversy, we explore two cases:
$\phi_3$ free (Fit~1), or fixed (Fit~2) at $60^\circ$~\cite{CKM}.

\section{Results}

The fitted rates, $A_{\rm CP}$s, and especially
$A_{\pi\pi}$ and $S_{\pi\pi}$, together with inputs,
are given in Table~\ref{tab:output}.
Setting $\delta$ and $\sigma$ to zero but keeping
all other parameters as determined by the fit,
the results are given in parentheses to indicate FSI cross-feed.
Note that $\sigma$ appears only in the
$\pi^-\pi^+$, $\pi^0\pi^0$, $K^-K^+$, $K^0\overline K^0$,
$\pi^0\eta_8$ and $\eta_8\eta_8$ rescattering subset.
The $\chi^2_{\rm min.}$ and fitted parameters are given in
Table~\ref{tab:phase}.
The $\chi^2_{\rm min.}/{\rm d.o.f.}$ for Fit 1 and 2 are 17/5
 (giving $\phi_3 \cong 96^\circ$)
vs. 25/6, and the former seems better. Both fits are much worse
without FSI: $\chi^2_{\rm min.}/{\rm d.o.f.}$ is 50/7 (65/8) for
Fit~1 (2), as seen in the last column of Table~\ref{tab:phase}.
For illustration, we obtain output errors for both Tables by scanning the
$\chi^2\leq \chi^2_{\rm min.}+1$ parameter space.

\begin{table}[b!]
\caption{
The $\chi^2_{\rm min.}$ and fitted parameters for Fits 1, 2.
We  constrain $f_\pi F_0^{BK}/f_K F_0^{B\pi} = 0.9\pm 0.1$,
and $1/m_s^{\rm eff}$ gives the effective chiral enhancement for
$\langle O_6\rangle$~\cite{AKL}.
The last column is for $\phi_3$ free (fixed) without FSI phases.
 \label{tab:phase} }
\begin{ruledtabular}
\begin{tabular}{crrc}
      &Fit 1
      &Fit 2
      &No FSI
      \\
\hline
 $\chi^2_{\rm min.}/{\rm d.o.f.}$
        & $\,\,17/5$
        & $\,\,25/6$
        &$50/7$ ($65/8$)
       \\
 $\phi_3$
        & \ $(96^{+21}_{-10})^\circ$
        & \ $[\,60^\circ]$
        &$106^\circ$ ($[\,60^\circ]$)
       \\
 $\delta$
        &  $(67^{+21}_{-11})^\circ$
        &  $(63^{+6}_{-0})^\circ$
        & ---
        \\
 $\sigma$
        & $(90^{+14}_{-59})^\circ$
        &$(103^{+0}_{-4})^\circ$
        & ---
        \\
 $F_0^{B\pi}$
        & $0.29^{+0.04}_{-0.02}$
        & $0.24^{+0.00}_{-0.01}$
        &0.25 (0.16)
    \\
 $F_0^{BK}$
        & $0.33^{+0.05}_{-0.06}$
        & $0.27^{+0.00}_{-0.02}$
        &0.27 (0.18)
    \\
 $m_s^{\rm eff}$ (MeV)
        & $81^{+26}_{-14}$
        &  $57^{+1}_{-3}$
        & 66 (35)
    \\
\end{tabular}
\end{ruledtabular}
\end{table}

\begin{figure}[t!]
\centerline{
    {\epsfxsize1.75 in \epsffile{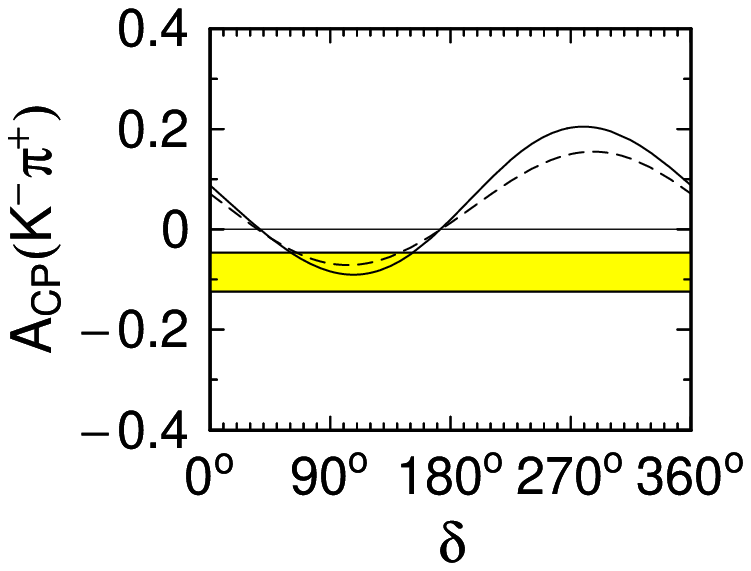}}
\hskip-0.1cm
        {\epsfxsize1.75 in \epsffile{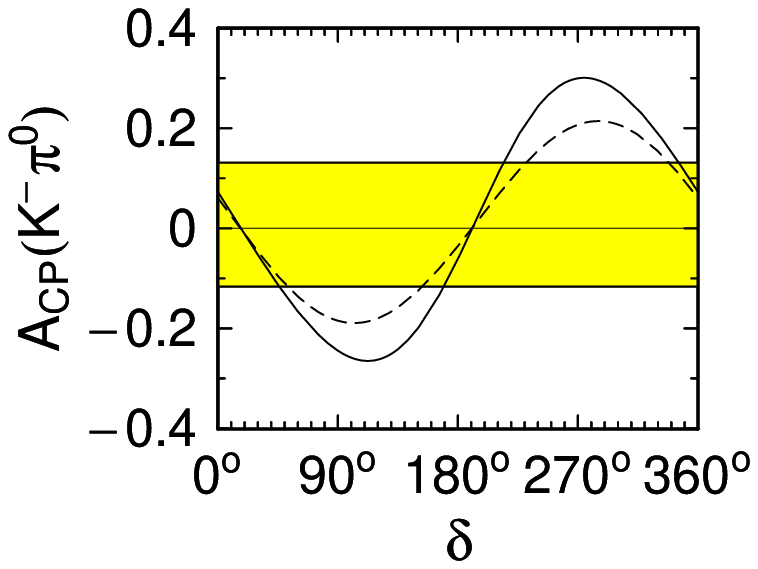}}}
\caption{ $A_{\rm CP}(K^-\pi^{+,0})$ vs. $\delta$.
Solid (dashed) line is for Fit~1 (Fit~2), and
shaded bands are 1$\sigma$ experimental ranges. }
\label{fig:acp}
\end{figure}

Let us illustrate the roles played by various inputs. The $K\pi$
rates are now measured with some precision. As before, together
with the large $K^-\pi^+/\pi^-\pi^+$ ratio, these are the main
driving force for large $\phi_3$~\cite{gamma,HSW}. What is new is
the pull of $A_{\rm CP}$s.
With the 3$\sigma$ effect from Belle gone, $A_{\rm CP}(K^0\pi^-)$
is consistent with zero and not very constraining. But the $A_{\rm
CP}$s in $K^-\pi^+$ now has some significance ($\sim 2\,\sigma$),
with the central value {\it opposite in sign} with respect to
factorization, which may call for~\cite{delta} rescattering. We
illustrate in Fig.~\ref{fig:acp} the $\delta$ dependence of
$A_{\rm CP}(K^-\pi^{+,0})$. Indeed, we see that for both fits, a
finite, positive $\sin\delta$ can turn the two $A_{\rm CP}$s
negative. Note that BaBar and CLEO give a negative $A_{\rm
CP}(K^-\pi^0)$, while the present Belle result flips sign from its
previous result and prefers a positive asymmetry. Although
$\sin\delta<0$ is allowed by rate data, it is disfavored by
$A_{\rm CP}(K^-\pi^+)$.
The two modes compete and settle on the fit output of $\delta \sim
65^\circ$, i.e. the $A_{\rm CP}(K^-\pi^+)$ preference for larger
$\delta$ is held back by $A_{\rm CP}(K^-\pi^0)$. A similar tug of
war is seen between the rates of $\overline K{}^0\pi^0$ (and
$\pi^-\pi^+$) vs. $K^-\pi^{+,0}$, $\overline K{}^0\pi^-$.

\begin{figure}[b!]
\centerline{
    {\epsfxsize1.67in \epsffile{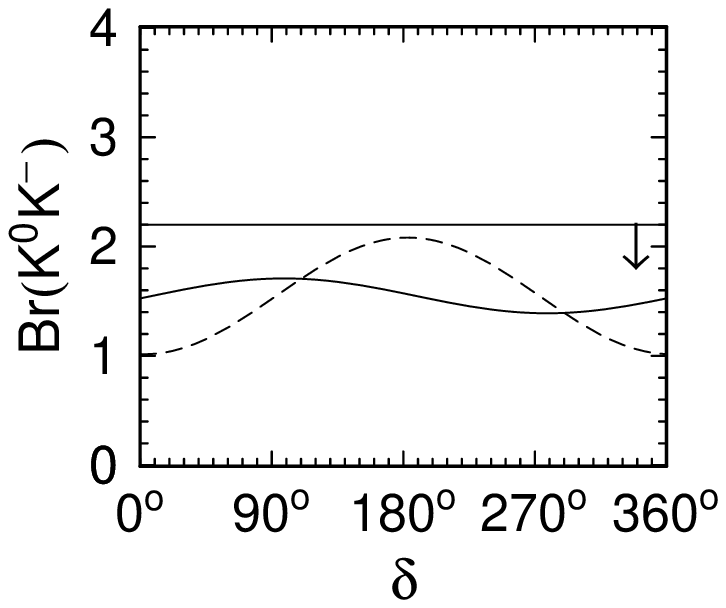}}
\hskip-0.2cm
        {\epsfxsize1.67in \epsffile{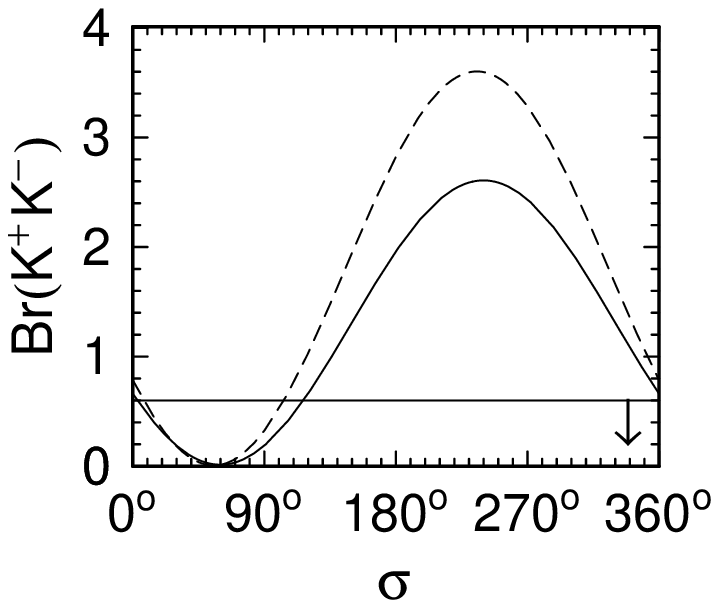}}}
\caption{ Rates ($\times 10^{6}$) for (a) $K^0K^-$ vs. $\delta$
and (b) $K^+K^-$ vs $\sigma$. Solid (dashed) line is for Fit 1
(Fit 2), with $\delta$ fixed at $55^\circ$ ($61^\circ$) in (b).
See text for choice of experimental limits. } \label{fig:kk0kk}
\end{figure}

As remarked earlier, the stringent bounds on $K^-K^{0,+}$ rates
require special care. The limits are not Gaussian so should not
enter the $\chi^2$ fit, but were enforced as strict bounds. As can
be seen from Table~\ref{tab:output}, while the fitted $K^-K^0$
rate is below the bound, for $K^-K^+$ the Fit~2 output sits right
at the bound. For Fit~1 it is also rather close. This implies
sensitivity to the bounds and the way they are implemented.
In Fig.~\ref{fig:kk0kk} we give the $\delta$ ($\sigma$) dependence
of $K^-K^{0(+)}$ and the bounds that we employ. Note that a
vanishing $K^+K^-$ rate is possible with finite $\sigma$. We
stress that data is still fluctuating, as both BaBar and Belle
have raised their bounds on $K^-K^0$ rates from the summer 2002
results.

Unlike $K^0K^-$ mode which is already $\sim 10^{-6}$ under
factorization, $K^-K^+$ is very suppressed hence sensitive to FSI.
Fig.~\ref{fig:kk0kk}(b) illustrates some subtlety of our SU(3) FSI
fit. From Eq. (\ref{eq:Smatrix}), if $\delta_{\bf 1}$,
$\delta_{\bf 8}$, $\delta_{\bf 27}$ are all randomly sizable, one
would expect $K^-K^+ \sim \pi^-\pi^+ > 10^{-6}$, which is realized
in Fig.~\ref{fig:kk0kk}(b) for $\vert \delta - \sigma \vert
\gtrsim 60^\circ$. But this is ruled out by the absence of
$K^-K^+$ so far. However, as also can be seen from
Fig.~\ref{fig:kk0kk}(b), for $\vert \delta - \sigma \vert \lesssim
50^\circ$, the $K^-K^+$ rate can be comfortably below the present
limit, {\it while $\delta$, $\sigma$ can be separately large}. The
subtlety is traced to the SU(3) decomposition
\begin{equation}
\left\langle (\pi\pi)_{I = 0} \left\vert {\cal S}^{1\over 2}
\right\vert (\pi\pi)_{I = 0}\right\rangle = \left( \frac{3}{8} \,
e^{i\delta_{\bf 1}} + \frac{3}{5} \, e^{i\delta_{\bf 8}} +
\frac{1}{40}e^{i\delta_{\bf 27}} \right).
\nonumber 
\end{equation}
Thus, because of the small weight of ${\bf 27} \to {\bf 27}$ in
the $I = 0$ $\pi\pi\to \pi\pi$ amplitude, when $\delta_{\bf 8}
\sim \delta_{\bf 1}$, i.e. $\delta \sim \sigma$, one finds $|
\langle (\pi\pi)_{I = 0}| {\cal S}^{1\over 2} \vert (\pi\pi)_{I =
0}\rangle \vert \simeq 1$, and ``leakage" to $\vert (K\overline
K)_{I = 0}\rangle$ is suppressed.
Note that $K^0\overline K{}^0 \sim K^0K^-$ also remains little perturbed.

\begin{figure}[t!]
\centerline{
            {\epsfxsize1.67 in \epsffile{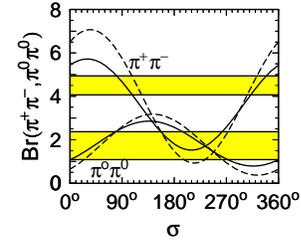}}
}
\caption { $\pi^+\pi^-$ and $\pi^0\pi^0$ rates ($\times 10^{6}$)
vs. $\sigma$. Solid (dashed) line is for Fit~1 (Fit~2) with
$\delta$ fixed at $67^\circ$ ($63^\circ$). Horizontal bands are
1$\sigma$ experimental ranges. } \label{fig:pipi}
\end{figure}

While $\sin\sigma < 0$ is strongly disfavored by $K^-K^+$ mode,
the driving force for large $\sigma$ rests in the $\pi\pi$ sector,
where {\it all} measurables turn out to be volatile
 --- {\it all} three rates and $A_{\pi\pi}$, $S_{\pi\pi}$.
The $\pi^-\pi^+$ and $\pi^0\pi^0$ rates are sensitive to both
$\delta$ and $\sigma = \delta_{\bf 27} - \delta_{\bf 1}$. To
illustrate, we fix $\delta$ to the fit values of
Table~\ref{tab:phase} and plot in Fig.~\ref{fig:pipi} the
$\pi^-\pi^+$, $\pi^0\pi^0$ rates vs. $\sigma$. The two fits are
similar and clearly favor large $\sigma \sim 90,\,103^\circ$, as
given in Table~\ref{tab:phase}.
The driving force for large $\sigma$ is the smallness of
$\pi^-\pi^+$ rate. For instance, if $\phi_3$ went upward by
$1\sigma$ in Fit~1, i.e. $\phi_3\sim 117^\circ$, the stress from
$\pi^-\pi^+$ rate would be greatly released and $\sigma$ can come
down to $\sim 31^\circ$.
But for Fit 2 where $\phi_3$ is held fixed at $60^\circ$,
one needs more FSI to reduce $\pi^-\pi^+$ rate by
shifting it partly to $\pi^0\pi^0$.
It is therefore intriguing that
{\it both Belle and BaBar have hints for the latter!}

At this point we note that the $\pi^-\pi^0$ rate in
Fit~2 is $\sim 3\sigma$ below experiment, and accounts for most of
the $\chi^2$ difference between Fits~1 and 2. Unless both
experiments are wrong, this makes Fit 2 less desirable; further
symptoms are the need for larger $1/m_s^{\rm eff}$ and lower
$F_0^{B\pi(K)}$. The situation comes about because, with low
$\phi_3$, it is hard to get low $\pi^-\pi^+$ rate, even with
rescattering. The fit resorts to reducing $F_0^{B\pi}$ by 20\%
(vs. Fit 1), hence reducing $\pi\pi$ rates by 36\%. Since
$\pi^-\pi^0$ rate is independent of $\phi_3$ and FSI,
this makes $\pi^-\pi^0$ too small.

\begin{figure}[t!]
\centerline{
            {\epsfxsize1.7 in \epsffile{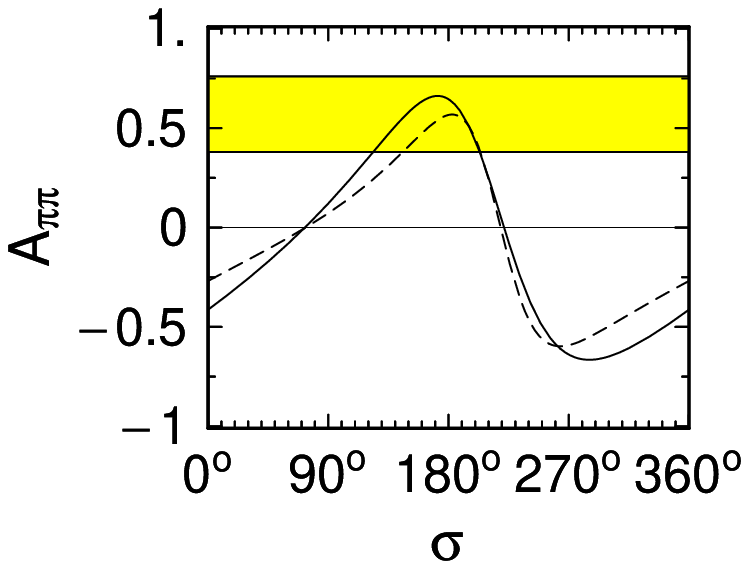}}
\hskip-0.25cm 
            {\epsfxsize1.7 in \epsffile{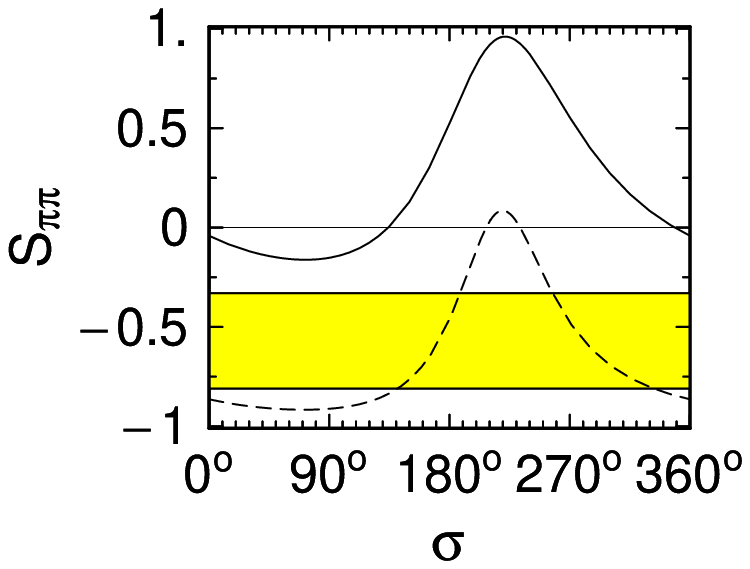}}
}
\caption { (a) $A_{\pi\pi}$ and (b) $S_{\pi\pi}$ vs. $\sigma$,
with notation as in Fig.~\ref{fig:pipi}. } \label{fig:apipispipi}
\end{figure}

What is intriguing --- and makes considering Fit~2 worthwhile
 --- is $A_{\pi\pi}$ and $S_{\pi\pi}$.
We plot these in Fig.~\ref{fig:apipispipi} with $\delta$ fixed at
fit values of Table~\ref{tab:phase}. While we are aware that the
two experiments are in conflict, we have forcefully ``combined"
the present Belle~\cite{pipi_belle} and BaBar~\cite{pipi_babar}
results (with enlarged error bars) in the plots. The experiments
do, however, agree on the {\it sign of $A_{\pi\pi}$}: Belle finds
$0.77\pm0.27\pm0.08$, and BaBar finds $0.30 \pm 0.25 \pm 0.04$
($\equiv -C_{\pi\pi}$). With $\sigma \simeq 100^\circ$, Fits 1 and
2 return $A_{\pi\pi}\simeq 0.12$, which agree with data in sign.
Without the FSI phases, both fits would give opposite sign.

As a measure of direct $CP$ violation, $A_{\pi\pi}$ is clearly
sensitive to FSI phases, but it does not distinguish much between
Fits~1 and 2; most results (except $\pi^-\pi^0$) are not that
different. It is $S_{\pi\pi}$, which probes the $CP$ phase of the
combined mixing and decay amplitudes, that is sensitive to
$\phi_3$, as is evident in Fig.~\ref{fig:apipispipi}(b). Various
$A_{\rm CP}$s and rates have constrained $\sin\delta > 0$ and
$\sin\sigma>0$. We have checked that $S_{\pi\pi}$ is not just flat
in $\sigma$ for $\sin\sigma > 0$ and $\delta \simeq 65^\circ$, as
seen from Fig.~\ref{fig:apipispipi}(b), but is relatively flat for
all $\delta,\ \sigma \lesssim 180^\circ$.
The sensitivity with $\phi_3$ gives rise to intriguing results.
{\it Fit~1} gives $\phi_3 \cong 96^\circ$, and $S_{\pi\pi} \sim
-15\%$ is {\it consistent with BaBar}. However, sensitivity to
$\phi_3$ brings $S_{\pi\pi}$ up to 50\% for $\phi_3=117^\circ$.
For {\it Fit~2}, one fixes $\phi_3 = 60^\circ$ to the CKM fit
value, and $S_{\pi\pi} \sim -0.9$ is {\it consistent with Belle}.
We have included Fit~2 in large part because of the present
volatility in $S_{\pi\pi}$. Otherwise it has much poorer $\chi^2$.

It is useful to compare with the
elastic (SU(2) or isospin) case given in Ref.~\cite{delta}.
For the whole range of $\delta$ and $|\delta-\sigma|\lesssim 70^\circ$,
the summed rates in $K^-\pi^+$--$\overline K{}^0\pi^0$
 ($\overline K{}^0\pi^-$--$K^-\pi^0$) and
 $\pi^-\pi^+$--$\pi^0\pi^0$ systems vary by
no more than few\% and 20\%, respectively.
We can reproduce the results of Ref.~\cite{delta}
for this parameter range by taking
$\delta_{K\pi} \equiv \delta_{3/2}-\delta_{1/2}
 \sim {\rm arg}\,(1+9 e^{i\delta})$
and $\delta_{\pi\pi} \equiv \delta_2 - \delta_0
 \sim {\rm arg}\,(1+24 e^{i\delta}+15 e^{i\sigma})$~\cite{erratum}.
We see from Fig.~\ref{fig:apipispipi}(a) that,
to reach the Belle and BaBar ``average" $A_{\pi\pi}$,
a very large $\sigma \sim 180^\circ$
 (hence large deviation from SU(2)) would be called for.
This situation is not supported by the absence of $K^-K^+$ and the
fact that $\pi^-\pi^+ > 4 \times 10^{-6}$.


\section{Discussion and Conclusion}

Some remarks are now in order.
First, the $S_{\pi\pi}$ sensitivity to $\phi_3$ and its numerics
are similar to other discussions~\cite{QCD,CKM,Fleischer:2002zv},
except we show that $S_{\pi\pi}$ is {\it insensitive} to FSI for $\sin\delta,\
\sin\sigma >0$.
Second, with current data, not only $\pi^0\pi^0 \gtrsim 10^{-6}$,
but also $K^-K^0$, $K^-K^+ \sim 10^{-6}$ are inevitable for our
rescattering model. Besides the rates of $\pi^-\pi^+$,
$\pi^0\pi^0$, they are largely driven by $A_{\rm CP}(K^-\pi^+)$
 (otherwise sign would be wrong).
It is also strongly suggested by
the need to change sign for $A_{\pi\pi}$.
With a factorization contribution at $10^{-6}$ level,
$K^-K^0$ should be seen soon, but
for $K^-K^+$, it is possible to get vanishing rates at
$\sigma\simeq\delta$, the ``SU(2) limit".
Third, our $\pi^-\eta_8$ rate is quite low and the $A_{\rm CP}$ is
opposite to the experimental result. This may be due to our
inability to treat $\eta_1$, or indicate an experimental problem,
or both.
Four, it has been pointed out that subleading, flavor exchange $PP
\to PP$ scattering need not be small at $m_B$
scale~\cite{Donoghue}. Note the present $\delta$ and $\sigma$
phases are {\it effective}~\cite{delta} parameters beyond
factorization. Their link to the actual $PP \to PP$ phases are
nontrivial and has to be studied further.
Fifth, a fully SU(3) analysis also hints at large FSI phases~\cite{He}.

Finally, let us compare with (P)QCD factorization approaches. For
QCD factorization, removing all subleading effects
 (hence $\sim$ naive factorization)
gives better fit~\cite{Beneke} to $K\pi$, $\pi\pi$ rates.
To account for $A_{\rm CP}$s, Ref.~\cite{QCD}
resorts to a sizable {\it complex} $X_{H,A}$ hadronic parameter,
which is a form of large strong phase.
PQCD factorization fares better with $A_{\rm CP}$s~\cite{keum} by
having a sizable absorptive part in some penguin annihilation
diagrams. But it gives a larger $\pi^-\pi^+$ and a smaller
$\pi^-\pi^0$ rate than data, the absence of $\pi^0\pi^0$, while
$K^-K^+$ is always very small~\cite{ChenLi}. Our results can
therefore be distinguished from QCD and PQCD factorization
approaches.

In summary,
we find that a minimal extension of
two quasi-elastic SU(3) FSI phases, $\delta$ and $\sigma$,
can suffice to account for current charmless $B$ decay data.
The observables to be tested in the near future are:
sign of $A_{\rm CP}$s, with $A_{\rm CP}(K^-\pi^{0}) \simeq -20\%$;
the rates of $\pi^0\pi^0 \gtrsim 10^{-6}$, $K^-K^+ \lesssim 10^{-6}$;
$A_{\pi\pi} > 0$, and $S_{\pi\pi}$ could, depending on $\phi_3$,
agree with either BaBar or Belle values.
The $\pi^-\pi^0$ rate should be double checked.
Many $A_{\rm CP}$s are large and have sign flipped, but can only
be checked later. More details of the present work would be given
elsewhere.

\noindent  
This work is supported in part by grants from NSC,
91-2112-M-002-027, 91-2811-M-002-043, 91-2112-M-033-013,
 the MOE CosPA Project,
and 
NCTS.


\begin{thebibliography}{99}
\bibitem{absorpt}
Short distance ``inelastic" phases via penguins
are included in standard way.
\end{thebibliography}
\end{document}